\documentclass[
article,
twocolumn,
english,
footinbib,
tightenlines,
nobibnotes,
aps,
prb,
showpacs,
floatfix
]{revtex4-2}
\usepackage{graphicx}
\usepackage[dvipsnames]{xcolor}
\definecolor{webblue}{rgb}{0,0,0.6}
\usepackage[
color links=true,
linkcolor=webblue,
citecolor=webblue,
urlcolor=webblue
]{hyperref}
\usepackage{amsmath}
\usepackage{amsfonts}
\usepackage{amssymb}
\usepackage{mathtools}
\usepackage{tikz}
\usepackage{bm}
\usepackage{bbm}
\usepackage{comment}
\usepackage[capitalize]{cleveref}
\usepackage{braket}
\usepackage[normalem]{ulem}

\definecolor{c-channel}{HTML}{488846}
\definecolor{p-channel}{HTML}{775271}
\definecolor{d-channel}{HTML}{AAA143}

\newcommand{\prlparagraph}[1]{\textit{#1}---}

\newcommand{\bvec}[1]{\boldsymbol{#1}}

\newcommand{\cre}{{\dag}}
\newcommand{\ann}{{\vphantom{\dag}}}

\usepackage{makerobust}
\newcommand{\makeauthor}[2]{\newcommand{#1}[1]{{%
  \protect%
  \color{#2}{%
    \bfseries\begingroup\escapechar=-1\edef\x{\endgroup\string#1}\x:%
  } ##1}}%
  \MakeRobustCommand#1}
\makeauthor{\matteo}{Plum}
\makeauthor{\lennart}{ForestGreen}
\makeauthor{\hendrik}{blue}
\definecolor{amaranth}{rgb}{0.9, 0.17, 0.31}
\makeauthor{\michi}{amaranth}
\makeauthor{\atanu}{orange}
\makeauthor{\jannis}{green}

\newcommand{\supplement}[1]{%
  \clearpage%
  \title{#1}%
  \maketitle%
  \setcounter{equation}{0}%
  \setcounter{figure}{0}%
  \setcounter{table}{0}%
  \setcounter{page}{1}%
  \makeatletter%
  \renewcommand{\thesection}{S\arabic{section}}%
  \renewcommand{\thesubsection}{\Alph{subsection}}%
  \renewcommand{\theequation}{S\arabic{equation}}%
  \renewcommand{\thefigure}{S\arabic{figure}}%
  \renewcommand{\thetable}{S\Roman{table}}%
  \renewcommand{\thepage}{S\arabic{page}}%
  \makeatother%
  \onecolumngrid%
}

\makeatletter
\def\maketitle{
\@author@finish
\title@column\titleblock@produce
\suppressfloats[t]}
\makeatother

\begin{document}


\title{Altermagnetic phase transition in a Lieb metal}

\author{Matteo D\"urrnagel}
\thanks{These authors contributed equally.}
\affiliation{Institut f\"{u}r Theoretische Physik und Astrophysik and W\"{u}rzburg-Dresden Cluster of Excellence ct.qmat, Universit\"{a}t W\"{u}rzburg, 97074 W\"{u}rzburg, Germany}
\author{Hendrik Hohmann}
\thanks{These authors contributed equally.}
\affiliation{Institut f\"{u}r Theoretische Physik und Astrophysik and W\"{u}rzburg-Dresden Cluster of Excellence ct.qmat, Universit\"{a}t W\"{u}rzburg, 97074 W\"{u}rzburg, Germany}
\author{Atanu Maity}
\affiliation{Institut f\"{u}r Theoretische Physik und Astrophysik and W\"{u}rzburg-Dresden Cluster of Excellence ct.qmat, Universit\"{a}t W\"{u}rzburg, 97074 W\"{u}rzburg, Germany}
\author{Jannis Seufert}
\affiliation{Institut f\"{u}r Theoretische Physik und Astrophysik and W\"{u}rzburg-Dresden Cluster of Excellence ct.qmat, Universit\"{a}t W\"{u}rzburg, 97074 W\"{u}rzburg, Germany}
\author{Michael Klett}
\affiliation{Institut f\"{u}r Theoretische Physik und Astrophysik and W\"{u}rzburg-Dresden Cluster of Excellence ct.qmat, Universit\"{a}t W\"{u}rzburg, 97074 W\"{u}rzburg, Germany}
\author{Lennart Klebl}
\affiliation{Institut f\"{u}r Theoretische Physik und Astrophysik and W\"{u}rzburg-Dresden Cluster of Excellence ct.qmat, Universit\"{a}t W\"{u}rzburg, 97074 W\"{u}rzburg, Germany}
\author{Ronny Thomale}
\email{ronny.thomale@uni-wuerzburg.de}
\affiliation{Institut f\"{u}r Theoretische Physik und Astrophysik and W\"{u}rzburg-Dresden Cluster of Excellence ct.qmat, Universit\"{a}t W\"{u}rzburg, 97074 W\"{u}rzburg, Germany}

\date{\today}


\begin{abstract}
We analyze the phase transition between a symmetric metallic parent state and itinerant altermagnetic order. The underlying mechanism we reveal in our microscopic model of electrons on a Lieb lattice does not involve orbital ordering, but derives from sublattice interference. 
\end{abstract}

\maketitle

\prlparagraph{Introduction}%
Triggered by the discovery of high-$T_c$ superconductivity in copper oxides, the possibility of unconventional pairing transcended our perspective on Fermi surface instabilities. Even though the individual two-electron ground state would always favor zero relative angular momentum according to the Perron-Frobenius theorem~\cite{Perron1907, Frobenius1912, Scammel2023, Ingham2024}, the condensation energy gain to the quantum fluid formed by phase-coherent Cooper pairs
drives
the preference of a $d$-wave particle-particle condensate, i.e., a pairing state with finite relative angular momentum. While the field has been aware of its principal possibility for decades, it has proven challenging to microscopically identify a similarly unconventional magnet, i.e., a spinful particle-hole condensate with finite relative angular momentum~\cite{Wu2007}. In combination with translation symmetry breaking, a rare instance is the kagome Hubbard model exhibiting a $p$-wave spin bond order~\cite{Kiesel2013}. In combination with point group symmetry breaking, a spin-type $d$-wave nematic order has been claimed in the context of higher order van-Hove singularities~\cite{Classen2020}, but has not yet unfolded in a generic microscopic model. The omnipresent difficulty in realizing such states is engraved in the oppositely charged constituents of the particle-hole pairs forming the quantum fluid, as they are naturally expected to gain condensation energy from zero relative angular momentum between particle and hole~\cite{Kiselev2017, Wu2018}.

A new chapter on broadening the phenomenological scope of exotic magnetic order has started with the discovery of a seemingly unprecedented type of magnetic order in 2019, now termed altermagnetism (AM)
~\cite{Smejkal2020,Naka2019,Ahn2019,Hayami2019,Yuan2020,10.1063/5.0005017,PhysRevLett.126.127701,Feng2022,PhysRevB.102.075112,PhysRevMaterials.5.014409,PhysRevB.102.144441,Ma2021, Marzin2021, Tamang2024}. The key driving interest rests upon the potential use of AMs for spintronics, as their energetically spin-split magnetic excitations, along with zero net magnetization, advantageously combine antiferromagnetic and ferromagnetic features for technological utilization~\cite{Gonzalez2021, Shao2021, Smejkal2022b, Guo2024}. A collinear AM is designed such that sites with opposite-spin orientation transform into each other by spatial rotation~\cite{Zeng2024, Cheong2024}. While this conceptually resembles the intertwining of time reversal and point group symmetry related to spin Pomeranchuk instabilities, a core insight embodied by the rising AM research domain is that such intertwining can just as well be more trivially achieved in a scale-separated cascade of transitions: A few proposals start from the formation of magnetic moments at high energy combined with a subsequent rotation symmetry breaking at lower energies~\cite{Brekke2023, Maeland2024, Bose2024, zhao2025altermagnetism}. As of now, the aspired abundance of AMs, however, mainly builds on the formation of anisotropies at the crystal field energy scale, followed by an independent magnetic transition determined by local moments and additional low-energy electronic scales~\cite{Smejkal2022, Smejkal2022a}.  A complete magnetic analogue of unconventional superconducting pairing, however, would only be resembled by the microscopic realization of a metallic parent state subject to a single phase transition into AM order~\cite{Pomeranchuk1958, Chubukov2018}.

In this Letter, we formulate a microscopic realization of an altermagnetic phase transition for interacting electrons on the Lieb lattice. Previous attempts to overcome the scale-separated nature of AM formation revolved around interacting electrons subject to combined staggered orbital and antiferromagnetic ordering. These proposals, however, are contrived in some aspects. In particular, the natural propensity of a staggered orbitally ordered system to form a ferromagnet due to Hund's coupling can only be avoided by ignoring the Hund's coupling altogether~\cite{Leeb2024,Giuli2024}. Instead, we seek to intertwine rotation and time-reversal symmetry through the sublattice interference (SI) profile found in a single-orbital Lieb metal parent state~\cite{Lin2024}.
Originally introduced for the kagome lattice~\cite{Kiesel2012,Kiesel2013,Jiang2022}, SI ascribes pivotal relevance to the sublattice distribution of Fermi level eigenstates with regard to identifying the relevant scattering channels for the unfolding electronic order. 
On the Lieb lattice, 
such interference leads to a dissociation of sublattice contributions to the magnetic fluctuation profile of the Lieb metal parent state. Through functional renormalization group (FRG) calculations~\cite{Salmhofer2001, Metzner2012, Platt2013, dupuis2021nonperturbative}, we show how this yields a phase transition into an AM phase. We classify the emerging order to descend from a $d$-wave spin Pomeranchuk instability, and to reveal a spin-split quasiparticle bandstructure featuring symmetry-protected nodal lines.

\begin{figure*}[]
    \centering
    \includegraphics[width=0.99\textwidth]{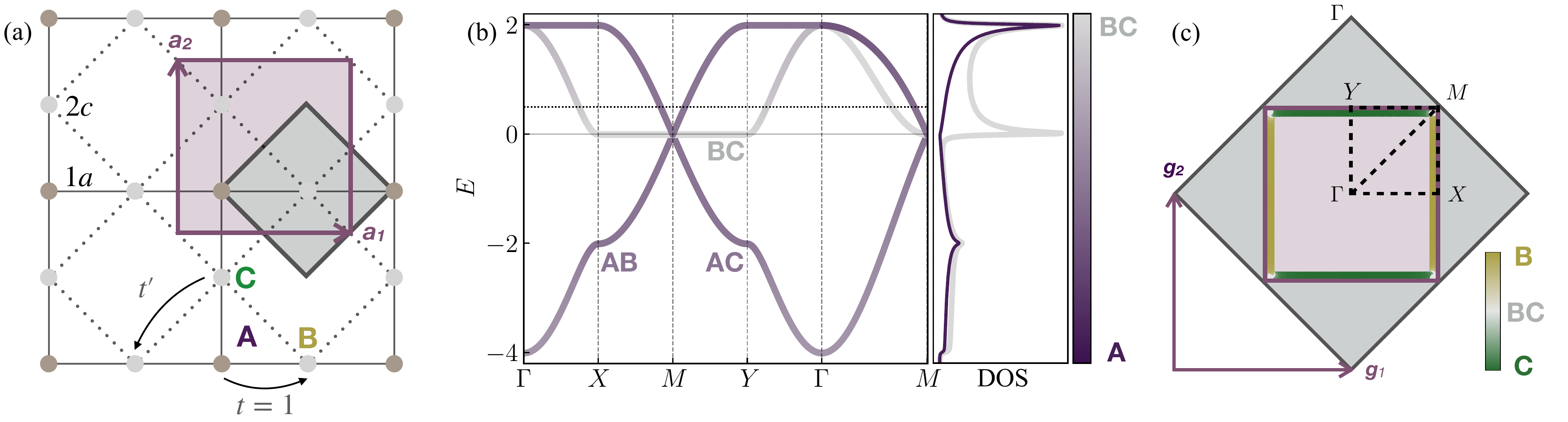}
    \caption{(a)~Real space lattice structure with of the Lieb lattice. The purple square marks the unit cell with the $A$ site on the trivial Wyckoff position ($1a$), and the $B$ and $C$ sites on the $2c$ Wyckoff positions. Nearest- and next-nearest neighbor hoppings $t$ and $t'$ are indicated. The gray square corresponds to the unit cell of a
    lattice without the $A$ site, i.e., a simple square lattice.
    (b)~Resulting band structure along the high-symmetry path indicated in panel~(c) for $t^\prime = t/2$ and $\mu_A=0$. We overlay the sublattice content by color (purple: $A$, gray: $B$/$C$). The right panel displays the sublattice resolved density of states, with the $B$/$C$ polarized van-Hove singularity at the band touching point.
    (c)~Brillouin zones (BZ) of the Lieb lattice structure (purple) and the intercalated square lattice (gray) 
    formed by dashed lines connecting the $2c$ Wyckoff positions. We additionally plot the Fermi surface and its sublattice polarization (yellow: $B$, green: $C$), which is pure along the BZ boundaries and mixed at the van-Hove points $M$.
    }
    \label{fig:1}
\end{figure*}

\prlparagraph{Model}%
We start from tightly bound electrons on a Lieb lattice, which is a square-depleted 2D lattice with three sites per unit cell and features equivalent nearest neighbor bonds connecting inequivalent sites. While it is rarely the lattice structure that minimizes a generic crystal field potential, the Lieb lattice forms a central building block of (layered) perovskites~\cite{Tsai2015}, and as such is relevant for high-$T_c$ copper oxides when the oxygen sites are not integrated out, but kept for their low-energy description~\cite{Emery1987}. \Cref{fig:1}a illustrates the sublattices on Wyckoff positions $1a$ ($A$) and $2c$ ($B$, $C$). While the $A$ site transforms trivially under all elements of $C_{4v}$, the sites $B$ and $C$ map onto each other under $C_4$ rotations. This transformation behavior is inherited by the tight-binding Hamiltonian. It reads
\begin{equation}
    \label{eq:LiebHubbardModel}
    H = -\sum_{ij,\sigma} t^\ann_{ij} c^\cre_{i\sigma} c^\ann_{j\sigma}
    - \sum_{i,\sigma} \mu_in_{i\sigma}
    + U\sum_i n_{i\uparrow} n_{i\downarrow}\,,
\end{equation}
where the operator $c^{(\cre)}_{i\sigma}$ annihilates (creates) an electron with spin $\sigma$ on site $i$, $n_{i\sigma} = c^\cre_{i\sigma}c^\ann_{i\sigma}$ is the density operator, $\mu_i$ denotes the chemical potential at site $i$, and $U$ is the Hubbard onsite repulsion.
We employ an extended Lieb lattice model by setting  $t_{ij} = 1$ for nearest neighbors (NN, $A$-$B$ and $A$-$C$ bonds, solid lines in \cref{fig:1}a) and $t_{ij} = t'$ for next-nearest neighbors (NNN, $B$-$C$ bonds, dashed lines).
 To allow for the effect of different atomic species at Wyckoff positions $1a$ ($A$) and $2c$ ($B$, $C$), we include an intrinsic detuning $\mu_A \neq \mu_{B,C}$.

While the electronic spectrum of the metallic parent state displays the full space group symmetry, the multi-site unit cell exhibits a non-trivial transformation behaviour of the eigenstates (Bloch functions).
For the Lieb lattice with NN hopping only ($t^\prime = 0$), a flat band resides at the Fermi level featuring a distinct sublattice polarization on the zone boundary: Excluding the zone corner at $M$, only the $B$ and $C$ sites contribute to the electronic eigenstates (see SM~\cite{SM} for details)~\cite{Lin2024}.
While this feature persists in the presence of NNN hopping $t^\prime$, the flat band attains a sizable dispersion away from the zone boundary, producing a van-Hove singularity (VHS, see \cref{fig:1}b). At the van-Hove $M$ point, the system displays a topologically protected quadratic (triple) band touching for $\mu_A\neq 0$ ($\mu_A=0$)~\cite{Tsai2015}.
Even at pristine filling (dashed line in \cref{fig:1}b), the proximate VHS implies
a large density of states (DOS) participating in a tentative Fermi surface instability, rendering the system already unstable against weak repulsive interactions.
The single-particle spectrum and sublattice polarization can be understood by decomposing the extended Lieb lattice into two intercalated square lattices. The $B$ and $C$ sublattice form a regular square lattice with lattice constant $1/\sqrt{2}$ and NN hopping $t^\prime$, which is coupled to the larger square lattice of the site $A$ via $t$ (cf. \cref{fig:1}a). Comparing the Brillouin zone (BZ) of the associated unit cells in \cref{fig:1}c, the Fermi surface (FS) at the zone boundary reveals itself as the characteristic FS of the smaller square lattice (gray) backfolded to the Lieb lattice BZ (purple).
While the weight of the $B$ and $C$ sublattices must be equal and constant in the the $t=0$ limit, the Lieb lattice structure allows for SI along the FS, manifesting in $B$/$C$ polarization along different directions.
Regardless of the value of $t'$, we find that the eigenstate along $X$-$M$ ($Y$-$M$) displays full $C$ ($B$) polarization. In turn, the high-energy VHSs display a strong $AB/AC$ mixing (see \cref{fig:1}b). This behavior is reminiscent of the pure/mixed VHSs found in the kagome lattice~\cite{Wu2021,Jiang2022,Hu2022,Kang2022}.

\begin{figure}
    \centering
    \includegraphics[width=\columnwidth]{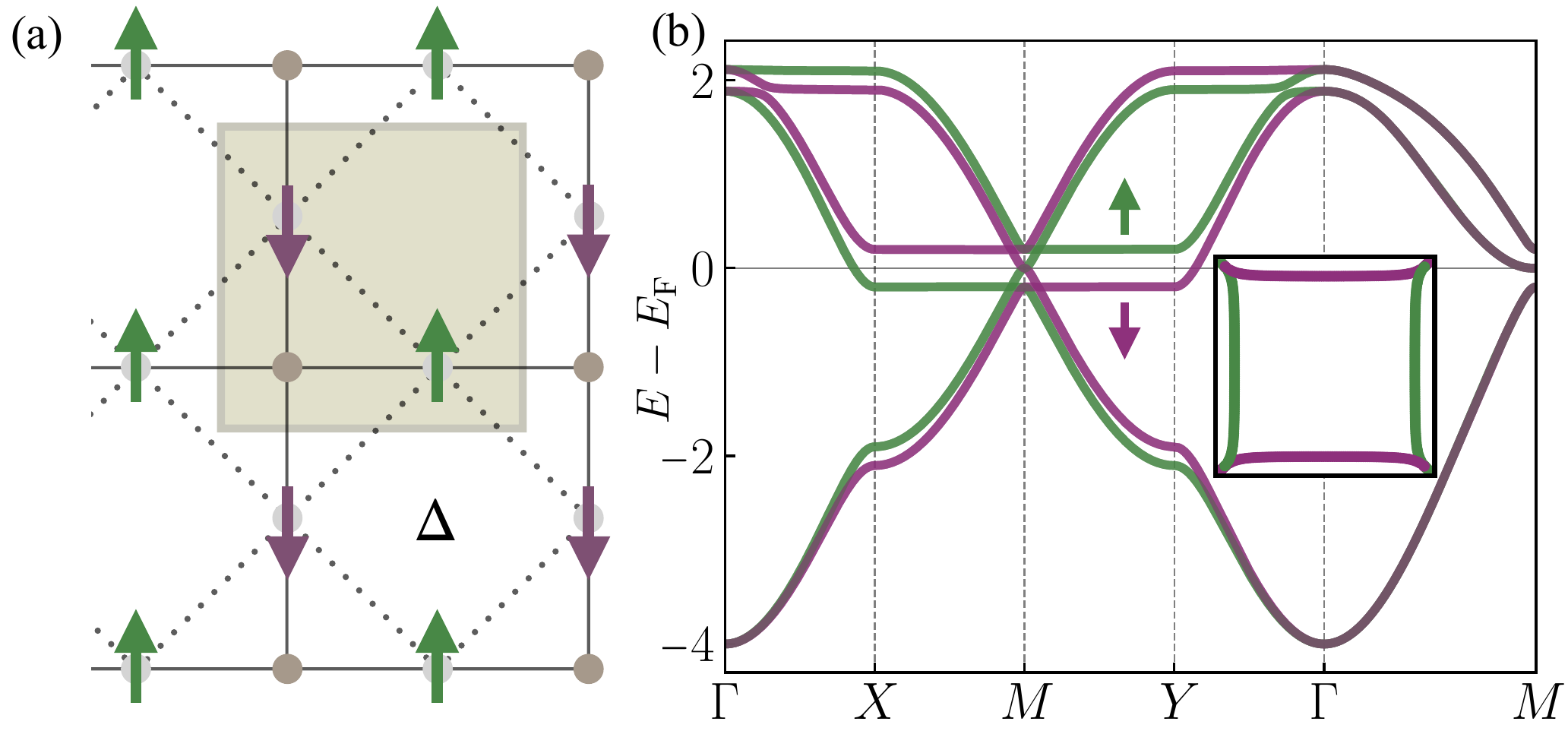}
    \caption{$d$-wave altermagnetic state on the Lieb lattice. (a)~Real space magnetization pattern in the $B_1$ irreducible representation of $C_{4v}$. The magnetic order parameter $\Delta$ is nonzero only on the $B$ and $C$ sites. (b)~Quasiparticle bandstructure in the altermagnetic phase for $t^{\prime}=t/2$, $\mu_A=0$, and $\Delta_{\mathrm{M}} =0.2\,t$ with spin polarization indicated by purple/green ($\downarrow$/$\uparrow$) lines. The inset shows the spin polarized Fermi surface, where the nontrivial transformation behavior under $C_4$ rotations and diagonal mirrors $\mathcal M_{xy}$, $\mathcal M_{x\bar{y}}$ becomes apparent.
    }
    \label{fig:2}
\end{figure}

\prlparagraph{AM order}%
Our Lieb metallic parent state implies an itinerant setup in which we investigate the unfolding of AM order. 
By contrast, as inspired by their \emph{ab initio} theoretical description, the prevalent perspective on AMs is the formation of local magnetic moments at high energy, and the subsequent low-energy magnetic ordering derives from antiferromagnetic exchange couplings of local moments in the given lattice geometry. This is why the symmetry classification of AMs deriving from local moments does not necessarily include all possible symmetry specifications of itinerant AMs~\cite{Roig2024}. Starting from a symmetric Lieb metal parent state, the ensemble of many-body phases competing with the AM \emph{a priori} has to be assumed to be vast. In particular, since both rotation symmetry and time-reversal symmetry jointly need to be broken to yield an AM transition, any charge-type nematic order or isotropic magnetic order has to be considered on equal footing in order to reach substantiated model evidence.
Therefore, an unbiased and general assessment of the many-body instabilities is required to avoid any \emph{a~priori} (mean-field) bias. Under these circumstances, the FRG in its optimized static four-point truncated unity approximation (TUFRG)~\cite{
husemann2009efficient, Lichtenstein2017, Profe2022, Beyer2022} is well-established to predict ordering propensities in an efficient manner~\cite{Profe2024a} (see SM~\cite{SM} for details).
Throughout the FRG flow, high-energy modes in all diagrammatic channels are integrated out and their \mbox{(anti-)} screening is incorporated into the resulting effective low-energy theory. Thereby, single-particle features like sublattice polarization and quantum geometry are imprinted into the interaction and may lead to non-trivial transformation behavior of the correlated many-body state under space group operations. We assume the kinetic energy scale, i.e., the electronic bandwidth $W$, to be the dominant energy scale over the interaction scale $U$, so that the perturbative diagrammatic resummation through the FRG flow equations is appropriate. Indeed, we checked that the FRG produces comparable results in a broad coupling range $U/t=0.1\dots3.8$ (see SM~\cite{SM}), as presented below.

In the Lieb lattice model (cf.~\cref{fig:1}a), we choose the Fermi level at the VHS close to half filling, expecting ordering tendencies to be strongest at points of high DOS. Given the FS geometry shown in \cref{fig:1}c, the FRG flows towards a magnetic instability irrespective of minor changes to the model~\cite{SM}. The renormalized interaction is dominated by commensurate particle-hole nesting contributions with transfer momentum $\bvec q = \bvec g_{1,2} \equiv \Gamma$, with $\bvec g_{1,2}$ reciprocal lattice vectors see \cref{sec:static-four-point-tufrg} of the SM~\cite{SM} for details on interpreting TUFRG results).
Such ``ferromagnetic''{} (disregarding sublattice structure) fluctuations have previously been ascribed to the Lieb lattice's non-trivial quantum-geometry~\cite{kitamura2024spin}, which is captured by TUFRG. \Cref{fig:2}a presents a real-space picture of the magnetic order parameter $\Delta$, which has been found to be stable at the mean-field level~\cite{Tsai2015}. In momentum space, its $d$-wave transformation behavior ($B_1$ irreducible representation of $C_{4v}$) becomes apparent, i.e.,
\begin{equation}
    \hat{\Delta}(\bvec k) = \hat{\sigma}_z \Delta_M \big[ \cos(k_x) - \cos(k_y) \big] \,,
\end{equation}
with $\Delta_M$ representing the magnitude and $\hat{\sigma}_z$ the Pauli-$z$ matrix in spin space. While the $B$/$C$ sites feature a sizable spin polarization, the $A$ site does not participate in the magnetic ordering process. The continuous deformation of the Fermi surface at the phase transition (cf.~\cref{fig:2}b) in the absence of translation symmetry breaking classifies the state as an $l=2$ spin Pomeranchuk instability (sPI)~\cite{Pomeranchuk1958, Chubukov2018}. Compared to Pomeranchuk's proposal, the Fermi momenta themselves do not exhibit a reduced symmetry; rather, only the spin polarization along the FS acquires a dependence on the Fermi momentum. This implies TRS breaking without Kramers degeneracy, and provides the non-relativistic $P$-$2$ spin momentum locking characteristic for $d$-wave AMs~\cite{Smejkal2022a}.

In the Lieb lattice structure, the absence of a finite magnetisation on the $A$ site is tightly related to the emergence of an AM state~\cite{Roig2024}. If a site on the trivial Wyckoff position $1a$ contributes to the magnetic order, the intra-unit cell structure of the magnetic ordering vector is bound to the trivial $A_1$ representation, and a spin-compensated magnetic state can thus only arise from broken translation symmetry, which would result in an AFM-type state. Intuitively, a finite magnetisation on the $A$ site might be expected due to the higher coordination of the $1a$ Wyckoff position, and hence the resulting increased energy gain in the ordered state. Instead, the itinerant nature of the present instability counteracts a gap on the $A$ site by SI. In particular around van-Hove filling, the scattering channels involve only the $B$ and $C$ sites. Notably, magnetization patterns involving the $A$ site are suppressed for $t > t'$ (i.e., the materials-oriented setting, see SM~\cite{SM}), promoting AM order on the $2c$ positions~\footnote{In the limit $t'=0$, Lieb's theorem holds, and we expect a finite magnetization. Setting $t'\neq0$, however, does not allow for the same conclusion, enabling AM order to emerge.}. In more general terms, the emergence of the AM state is tied to the sublattice structure at the VHS, which persists even in the case of longer-ranged hoppings and interactions~\cite{SM}.
Our results highlight that in the itinerant picture, magnetic ordering on the $A$ site is not suppressed \emph{a~priori} by crystal field effects. Instead, the precise nature of the states at the Fermi level enables SI to percolate into the renormalized interaction, and eventually to favor an AM instability.

\begin{figure}
    \centering
    \includegraphics[width=0.95\columnwidth]{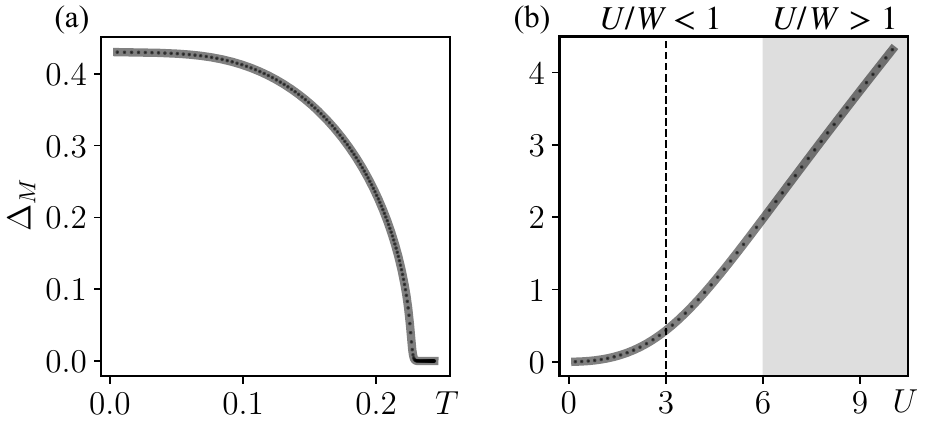}
    \caption{(a)~Size of the altermagnetic order parameter ($\Delta_M$) as a function of temperature ($T$) for $t'=t/2$ and $U=3t$, obtained from self-consistent mean-field simulations, showing a clear mean-field phase transition at $T_c/t \approx 0.23$. (b)~Value of the magnetic order parameter in the limit of $T=0$ for $U$ across a large parameter regime.
    The dashed line indicates $U=3t$ as used in panel~(a).
    }
    \label{fig:3}
\end{figure}

\Cref{fig:2}b demonstrates how the quasiparticle band structure inherits the non-trivial transformation behavior of the $2c$ sites under $C_{4}$ rotations and diagonal mirrors $\mathcal M_{xy}$, $\mathcal M_{x\bar{y}}$:
In addition to the non-relativistic spin-splitting along $\Gamma$-$X$ ($\Gamma$-$Y$), the magnetic gap $\Delta(\bvec k)$ exhibits symmetry-protected nodal lines along $\Gamma$-$M$. 
To gain insights on the AM phase transition, we use our FRG result to perform a constrained self-consistent mean-field analysis (see \cref{sec:mean_field} of the SM~\cite{SM} and Refs.~\cite{Wang2014, O2024}). We find a second-order transition with $\Delta$ smoothly increasing from $T = T_c$ to its maximum value at $T=0$ (cf.~\cref{fig:3}a). At the VHS, where the divergent DOS leads to the largest induced spin gap, we demonstrate that significant values of $\Delta$ can be reached: \Cref{fig:3}b displays the mean-field result for $\Delta_M$ as a function of $U$ at zero temperature.
Even in the weak to intermediate coupling regime ($U/W < 1$, white region), $\Delta_M$ is on the order of $t$. In turn, the size of the spin splitting $\delta$ is readily given by the magnitude of the altermagnetic order parameter $\Delta_M$ for $\Delta_M \ll t$~\cite{SM} to read
\begin{equation}
    \delta\vert_{\bvec k=X} = \begin{cases}
        2\Delta_M & \quad \text{for the central band,} \\
        \Delta_M \, |1\pm \mu_A/t| & \quad \text{for the outer bands.}
    \end{cases}
\end{equation}
Facing current insecurities about how a theoretically anticipated AM spin splitting scale might relate to the actual splitting scale resolved in experiment, this result encourages the conjecture that AM order resulting from an AM phase trasition might not share the big crystal field anisotropy scales, but could still be competitive or even suprerior in terms of the reachable AM spin splitting.

\prlparagraph{Discussion}%
We show how an itinerant electronic Fermi surface instability mechanism can lead to an altermagnetic phase transition. Instead of the conventional hierarchy in altermagnetic materials, where a high-energy crystal field effect precedes magnetic ordering at lower energy scales, our approach makes use of the 
sublattice texture imprinted into electronic correlations through the Fermi level. As a consequence, only some of the degrees of freedom are selected in the formation of a magnetic state, leading up to a spin Pomeranchuk instability. In the ordered phase, the quasiparticle band structure displays features known from altermagnets, i.e., non-relativistic spin splitting with net zero magnetization in the absence of translation symmetry breaking. We substantiate the itinerant mechanism at the example of a Hubbard model on the Lieb lattice. By means of the FRG, we obtain its weak coupling instabilities and find a prominent tendency towards an $l=2$ ($d$-wave) spin Pomeranchuk order. Notably, the crystal structure does not fall into the classification of altermagnets~\cite{Guo2023, Wei2024, Smolyanyuk2024, Roig2024}. Instead, sublattice interference leads to selective site participation in the magnetic order. Our obtained magnetization pattern then is equivalent to the $P$-$2$ altermagnetic state featured in toy models of RuO\textsubscript{2}~\cite{Smejkal2022,Roig2024}. 

The absence of altermagnetic signatures in RuO\textsubscript{2}~\cite{liu2024absence, plouff2024revisiting} as well as MnF\textsubscript{2}~\cite{morano2024absence} candidate materials is remarkable in face of an expectation of large AM energy scale from mean field theory. This might point to a substantial screening effect so far overlooked in the hierarchical, scale-separated symmetry breaking mechanism of local moment AM setups, deserving further investigation. In addition, Ref.~\cite{Wan2024} suggests that altermagnetic \emph{metals} are rather challenging to find assuming hierarchical symmetry breaking.
In this context, the lack of scale separation embodied by our AM mechanism establishes a notable difference, which might bring about beneficial conditions for AM order and its descendant spin splitting. We expect our itinerant AM formation principle to unlock new altermagnetic states in crystallographic geometries outside the conventional classification of altermagnets presented in Refs.~\cite{Guo2023, Wei2024, Smolyanyuk2024, Roig2024},
given the abundance of sublattice interference in various two- and three-dimensional lattices~\cite{Lin2024}.
Since our AM mechanism offers a direct transition from a metallic to an altermagnetic state, it can be probed by, e.g., spin-dependent scanning tunneling microscopy or angle-resolved photoemission spectroscopy. Some candidate systems include covalent organic systems~\cite{Cui2020}, metal-organic frameworks~\cite{Che2024}, and optical lattices (see SM~\cite{SM} for the Lieb lattice, and Ref.~\cite{das2024realizing} for a square lattice proposal). Furthermore, the approach presented here is not restricted to collinear AM order, but can likewise be applied in the context of non-collinear altermagnetism~\cite{Cheong2024a}.

\prlparagraph{Note added}%
Upon completion of this work, we became aware of an independent work by Li~\emph{et.~al.}~\cite{Li2024}, which explores altermagnetism in the Emery model for cuprates. Their and our model relate to each other in the extreme parameter regime $\mu_A \ll 0$.

\medskip

\prlparagraph{Acknowledgments}%
We thank R.~Samajdar and T. M\"uller for useful discussions.
M.~D.~thanks the group of S.~Huber at ETH Zürich for their hospitality.
This work is supported by the Deutsche Forschungsgemeinschaft (DFG, German Research Foundation) through Project-ID 258499086 -- SFB 1170 and through the W\"urzburg-Dresden Cluster of Excellence on Complexity and Topology in Quantum Matter -- ct.qmat Project-ID 390858490 -- EXC 2147.
We are grateful for HPC resources provided by the Erlangen National High Performance Computing Center (NHR@FAU) of the Friedrich-Alexander-Universit\"at Erlangen-N\"urnberg (FAU), that were used for the FRG calculations. NHR funding is provided by federal and Bavarian state authorities. NHR@FAU hardware is partially funded by the DFG -- 440719683. 

\let\oldaddcontentsline\addcontentsline
\renewcommand{\addcontentsline}[3]{}
\bibliography{references.bib}
\let\addcontentsline\oldaddcontentsline

\supplement{Supplementary material:\\
Altermagnetic phase transition in a Lieb metal}

\tableofcontents

\section{Sublattice content of the low-energy van-Hove singularity}
Transforming the kinetic part of the Hamiltonian [cf.~\cref{eq:LiebHubbardModel}] to momentum space, we obtain a matrix in sublattice space ($A,B,C$) for each BZ momentum $\bvec k$:
\begin{equation}
    H(\bvec k) = \begin{pmatrix}
        - \mu_A & A_{k_x} & A_{k_y} \\
        A_{k_x} & 0 & B_{\bvec k} \\
        A_{k_y} & B_{\bvec k} & 0
    \end{pmatrix} \,.
    \label{eq:sm_ham}    
\end{equation}
Here, we dropped its spin dependence due to $SU(2)$ symmetry. Without loss of generality, we set $\mu_B = \mu_C = 0$. The coefficients of \cref{eq:sm_ham} are given as
\begin{equation}
    \label{eq:sm_ham_coeff}
    A_{k_{x}} = 2t \cos\Big(\frac{k_x}{2}\Big) \,, \qquad
    A_{k_{y}} = 2t \cos\Big(\frac{k_x}{2}\Big) \,, \qquad
    B_{\bvec k} = 4t'\cos\Big(\frac{k_x}{2}\Big) \cos\Big(\frac{k_y}{2}\Big) \,.
\end{equation}
From the functional form of \cref{eq:sm_ham,eq:sm_ham_coeff} is it directly evident that a zero energy state persists at the zone boundary (in the presence of arbitrary $t^\prime$) since there, $\cos(k_x/2) = 0 \, \lor \, \cos(k_y/2) = 0$. Along a zone boundary path, e.g., $M$-$X$, the Hamiltonian acquires the simple form
\begin{equation}
    H(\bvec k) =
    \begin{pmatrix}
        - \mu_A & 0 & 2t \cos(k_y/2) \\
        0 & 0 & 0 \\
        2t \cos(k_y/2) & 0 & 0
    \end{pmatrix} \,,
    \label{eq:sm_ham_edge}
\end{equation}
and the polarization of the zero mode eigenstate in the $B$ sublattice is directly apparent from the matrix form of \cref{eq:sm_ham_edge}. At $M = (\pi, \pi)$, this is a zero triple state for $\mu_A = 0$ and evolves into a quadratic band touching point at finite $\mu_A$ as the Dirac cone is gapped out by the Semenov-like mass term in analogy to the Honeycomb lattice~\cite{Tsai2015}.

Adding additional hoppings beyond second nearest neighbor (e.g.~diagonal $B$-$B$ and $C$-$C$ hoppings across the plaquettes) leads to a warping of the Fermi surface in correspondence to the square lattice Hubbard model with NNN hoppings. While this reduces the nesting features of the FS and leads to a reduction of the critical temperature, the orbital character and nesting vector of the VHS persists and continues to support the altermagnetic phase.

\section{Effective model for the $B$/$C$ sites}
\label{sec:perturbation_theory}
Starting with the Hamiltonian defined in \cref{eq:sm_ham}, we can write it as a block matrix in sublattice space:
\begin{equation}
    H = \begin{pmatrix}
        H_A & T^\dagger \\
        T & H_{BC}
    \end{pmatrix} \,.
\end{equation}
The corresponding free Greens functions for the two subspaces are given by $\mathcal G_{A,BC}^0 = (i\omega - H_{A,BC})^{-1}$. We build an effective model for the $B$ and $C$ sublattices by treating $T\propto t$ as perturbation. Note that in our model, however, $t$ is not a small parameter and therefore a perturbative treatment is not justified \emph{a priori}. The effective $BC$ subspace Green's function then reads~\cite{karrasch2010functional, Kalkstein1971, LopezSancho1985}
\begin{equation}
    \def\GAF{{{\mathcal G}_{A}^{0}}}
    \def\GBCF{{\big({\mathcal G}_{BC}^{0}\big)}}
    \mathcal G_{BC} = \frac1{\GBCF^{-1} - T \GAF T^\dagger} ={}
    \frac{1}{i\omega - H_{BC} - T\GAF T^\dagger}
    \equiv
    \frac1{i\omega - H_{BC} - \Sigma_{BC}}
    \,.
\end{equation}
The (hybridization) self-energy $\Sigma_{BC}$ is therefore given as
\begin{equation}
    \Sigma_{BC}(i\omega, \bvec k) = \frac{4t^2}{i\omega + \mu_A} \begin{pmatrix}
        \cos^2(k_x/2) & \cos(k_x/2)\cos(k_y/2) \\
        \cos(k_x/2)\cos(k_y/2) & \cos^2(k_y/2)
    \end{pmatrix} \,,
\end{equation}
which we can recast into an effective Hamiltonian in the zero frequency limit:
\begin{equation}
    H^\mathrm{eff}_{BC}(\bvec k) = H_{BC}(\bvec k) + \Sigma_{BC}(0,\bvec k) = \begin{pmatrix}
        4 \mathsf Z \cos^2(k_x/2) & 4(t' + \mathsf Z) \cos(k_x/2)\cos(k_y/2) \\
        4(t' + \mathsf Z) \cos(k_x/2)\cos(k_y/2) & 4 \mathsf Z \cos^2(k_y/2) \\
    \end{pmatrix} \,,
\end{equation}
where we defined the expansion parameter $\mathsf Z = t^2/\mu_A$ for brevity.
Rotating the coordinate system by $45^\circ$ ($2q_x = k_x + k_y$, $2q_y = k_y - k_x$) aligns the BC plaquette with the $x$-$y$ directions, which results in
\begin{equation}
    H^\mathrm{eff}_{BC}(\bvec q) = \begin{pmatrix}
        2\mathsf Z \big(1+\cos(q_x/2 + q_y/2)\big) & (t' + \mathsf Z) \big(\cos(q_x/2) + \cos(q_y/2)\big) \\
        (t' + \mathsf Z) \big(\cos(q_x/2) + \cos(q_y/2)\big) &
        2\mathsf Z \big(1+\cos(q_x/2 - q_y/2)\big)
    \end{pmatrix} \,.
\end{equation}
The second order processes in $t$ (first order in $\mathsf Z$) breaks the local $C_4$ symmetry around the $B$ and $C$ sites by introducing direction dependent hoppings on the Hamiltonian's diagonal. This scenario is reminiscent of the effective models for AM in RuO\textsubscript{2} (compare e.g. Eq.~(17) in Ref.~\cite{Roig2024}).

\section{Details on the (TU)FRG calculations}

\subsection{Static four-point TUFRG}
\label{sec:static-four-point-tufrg}

\begin{figure}
    \centering
    \includegraphics{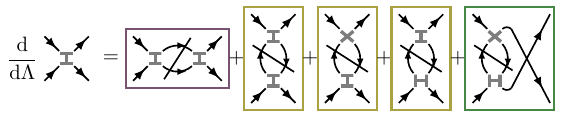}
    \caption{Diagrammatic representation of static four-point FRG. We group the three channels by color: particle-particle channel ({\color{p-channel}$P$, purple}), direct particle-hole channel ({\color{d-channel}$D$, yellow}), and crossed particle-hole channel ({\color{c-channel}$C$, green}). Spin is conserved along the short edge of each vertex.}
    \label{fig:diagrams}
\end{figure}
We use the FRG in its static four-point approximation. This means that we disregard all higher (six, eight, \dots) point interaction vertices as well as self-energies in the expansion of the vertex generating functional~\cite{Metzner2012}. The resulting diagrammatic structure of the FRG flow equations is visualized in \cref{fig:diagrams}. In addition to the trunctation of the number of fermionic fields participating in possible interactions, we disregard the frequency dependence of all vertices. This approximation has proven useful when characterizing ordering propensities of weakly to intermediately interacting electron systems~\cite{Salmhofer2001}. As the four-point vertices are still complicated objects, they are routinely compressed in numerical simulations: The primary (bosonic) momentum is resolved in momentum space, while the secondary (ferminoic) momenta are treated in a truncated real-space basis resulting in three distinct diagrammatic channels (particle-particle ($P$), crossed particle-hole ($C$), and direct particle-hole ($D$)). This coins the ``truncated unity''{} approximation of FRG, i.e., TUFRG~\cite{husemann2009efficient, Lichtenstein2017, Profe2022, Beyer2022}.

The renormalization group flow is terminated upon encountering a divergent vertex element at some critical cutoff $\Lambda_c$ indicating an instability of the Fermi liquid towards a symmetry broken phase.
Since all quantum fluctuations above $\Lambda_C$ are included in the effective vertex, the residual effective interaction can be analyzed within a simple MF theory to reveal the expected symmetry breaking~\cite{Reiss2007, Platt2013}.
In order to carry out the Hubbard-Stratonovich transformation of the effective 4-point vertex, a fermionic bilinear has to be chosen as static MF.
There are three possible choices, corresponding to superconducting
($\langle c^\dagger_{\boldsymbol{k}', \uparrow} \, c^\dagger_{\boldsymbol{q} - \boldsymbol{k}', \downarrow} \rangle$),
charge ($\langle c^\dagger_{\boldsymbol{k}', \uparrow} \, c^{\phantom{\dagger}}_{\boldsymbol{q} + \boldsymbol{k}', \uparrow} + c^\dagger_{\boldsymbol{k}', \downarrow} \, c^{\phantom{\dagger}}_{\boldsymbol{q} + \boldsymbol{k}', \downarrow} \rangle$)
or magnetic ($\langle c^\dagger_{\boldsymbol{k}', \uparrow} \, c^{\phantom{\dagger}}_{\boldsymbol{q} + \boldsymbol{k}', \uparrow} - c^\dagger_{\boldsymbol{k}', \downarrow} \, c^{\phantom{\dagger}}_{\boldsymbol{q} + \boldsymbol{k}', \downarrow}\rangle$) instability.
The MF interaction matrices $V$ for the respective orders can be directly obtained from the interaction in the diagrammatic channels depicted in \cref{fig:diagrams} as $V^\text{SC} = V^P$, $V^\text{charge} = 2 V^D - V^C$, and $V^\text{magnetic} = V^C$.

For the physical interaction channels a MF decoupling is performed by defining the non-vanishing expectation value \textit{i.e.} in the magnetic channel as
\begin{equation}
    \Delta_{\boldsymbol{q}, \,1\,4}(\boldsymbol{k}) \, = \, \sum_{\boldsymbol{k}', 3\,2} \, V^\text{magnetic}_{1\,4\,3\,2} (\boldsymbol{q}, \boldsymbol{k}, \boldsymbol{k}') \langle c^\dagger_{\boldsymbol{k}', \uparrow , 3} \, c^{\phantom{\dagger}}_{\boldsymbol{k}'+\boldsymbol{q}, \uparrow , 2} - \,
    c^\dagger_{\boldsymbol{k}', \downarrow , 3} \, c^{\phantom{\dagger}}_{\boldsymbol{k}'+\boldsymbol{q}, \downarrow , 2}
    \rangle,
\end{equation}
where all quantum numbers other than spin are combined in roman numeral multi-indices.
Since the FGR flow is only valid until the phase transition, it is meaningful to apply the MF theory directly at the phase boundary where $\Delta \rightarrow 0$.
In this limit the self-consistent gap equation ca be linearized and reduces to a simple eigenvalue equation, that ca be evaluated independently for every transfer momentum $\boldsymbol{q}$
\begin{equation}
    \lambda_{\boldsymbol{q}} \, \Delta_{\boldsymbol{q}, \zeta}\, =\, \sum_{\zeta'} V_{\zeta \, \zeta'}(\boldsymbol{q}) \, \Delta_{\boldsymbol{q}, \zeta'} \, ,
\end{equation}
where $\lambda_\mathbf q$ acts as a proxy for the transition temperature for the respective order.
The physically realized instability is then indicated by the eigenvector $\Delta$ associated with the largest eigenvalue $\lambda$ over all channels and wave vectors $\mathbf q$.
Here, the vertices are already represented in the matrix basis, absorbing the reordering complexity within the index $\zeta \leftrightarrow (1\,4\,\boldsymbol{k})$ and $\zeta' \leftrightarrow (3\,2\,\boldsymbol{k}')$.

\subsection{Numerical details}
The FRG calculations were performed with the TUFRG backend of the divERGe library, making use of the sharp frequency cutoff as it boosts numerical performance~\cite{Profe2024a}. We employed a $30 \times 30$ mesh for the bosonic momenta of the vertices, with an additional refinement of $51 \times 51$ for the integration of the loop. The form-factor cutoff distance is chosen as 2.51 in units of lattice vector length. We check for convergence by calculating selected points in parameter space with increased number of momentum points and
form-factor cutoff ($42\times 42$, refinement: $81 \times 81$, formfactor cutoff: 3.51). We employed the adaptive Euler integrator of
the divERGe library with default parameters. Calculations displayed in the main part of this work were obtained for nearest ($t=1$) and next-nearest neighbor hopping ($t'=0.5$), and equal onsite interactions and chemical potentials on the three sublattices ($U=1$, $\mu_i=0$) as given by the interaction Hamiltonian in \cref{eq:LiebHubbardModel} of the main text.

\subsection{Mean-field altermagnetic phase transition from FRG}
\label{sec:mean_field}
The truncation of the FRG flow equations breaks down when approaching a phase transition and results in a divergence of the two-particle vertex. Investigating the symmetry breaking beyond the phase transition predicted by FRG can be achieved by formulating an effective mean-field (MF) theory. To bridge the non-analytic behaviour of the FRG flow at the phase transition and smoothly connect the ordered state with the high temperature phase, we follow the procedure outlined in Refs.~\cite{Wang2014, O2024}: Close to the phase transition, the divergence stems from the ladder series of the diverging channel and is only marginally altered by cross channel projections. To obtain an appropriate bare interaction $V_\mathrm{MF}$ for MF treatment (at scale $\Lambda_\mathrm{MF} \gtrsim \Lambda_C$), one can extrapolate this regime to high energies. In practice, we `reverse'{} the ladder resummation in the divergent channel over the full range of renormalization group scales (here given in the magnetic channel):
\begin{equation}
    \Gamma^{\Lambda_\mathrm{MF}}(14,32) = V_\mathrm{MF}(14, 32) -
    \sum_{1'2'3'4'} V_\mathrm{MF}(14, 3'2') \, L(3'2',1'4') \, \Gamma^{\Lambda_\mathrm{MF}}(1'4', 32)
\end{equation}
with the appropriate bare susceptibility $L(12.34)$. Here, all quantum numbers are combined in roman numeral multi-indices. Inverting the above equation defines the starting point of the MF treatment. By construction, a subsequent MF analysis will yield the correct MF state.

In the present case, the AM symmetry breaking is directly triggered by the bare interaction: Even without cross-channel feedback, an RPA resummation in the magnetic channel yields the same AM instability (as evidenced by the pronounced peak in the particle-hole susceptibility spectrum displayed in \cref{Fig:sup3}). So the inversion of an RPA ladder outlined above results in the \emph{actual} bare $V_\mathrm{MF} \equiv U$. We can thus equivalently employ an MF decomposition of the bare Hamiltonian [cf.~\cref{eq:LiebHubbardModel}]. To derive the self-consistent MF equations we calculate the free energy at fixed total particle number $n_\mathrm{tot}$:
\begin{equation}
    F = - \frac{1}{\beta} \ln(\mathcal Z) + \mu n_\mathrm{tot} = - \frac{1}{\beta} \ln \Big( \int \mathcal{D}[\psi,\bar{\psi}] \, e^{-S(\psi, \bar{\psi})} \Big) + \mu n_\mathrm{tot} \,,
    \label{eq:free_energy}
\end{equation}
with the action $S$ given by
\begin{multline}
    S(\psi, \bar{\psi}) = S_0(\psi, \bar{\psi}) + S_I(\psi, \bar{\psi}) =
    \frac{1}{N} \sum_{k s s^\prime o_1 o_2} \bar{\psi}_{k o_1 s} (-i\omega_n \delta_{s s^\prime} \delta_{o_1 o_2} + H^0_{o_1 o_2}(\bvec k) \delta_{s s^\prime}) \, \psi_{k o_2 s^\prime} \\
    + \frac{U}{N^2} \sum_{k_i s s^\prime o} \bar{\psi}_{k_1 o s} \bar{\psi}_{k_2 o s^\prime} \psi_{k_3 o s^\prime} \psi_{k_1 + k_2 - k_3 o s} \,,
\end{multline}
where $k = (\bvec k, \omega_n)$ and $H^0(\bvec k)$ the non-interacting Hamiltonian given in \cref{eq:sm_ham}. We decouple the interaction with a Hubbard-Stratonovich transformation and constrain the bosonic fields to the static order parameter of the FRG calculation:
\begin{equation}
    \Delta_{o_1 o_2} = \frac{U}{N} \sum_{k s s^\prime} \langle \bar{\psi}_{k o_1 s} \sigma_z^{s s^\prime} \psi_{k o_2 s^\prime} \rangle
      = \begin{pmatrix}
            0 & 0 & 0 \\
            0 & \Delta_M & 0 \\
            0 & 0 & -\Delta_M 
        \end{pmatrix}\,,
    \label{eq:order_parameter}
\end{equation}
where the Pauli-$z$ matrix $\sigma_z$ fixes the magnetization axis without loss of generality. Neglecting fluctuations of around the MF state, we obtain a quadratic action that reads
\begin{equation}
    S(\psi, \bar{\psi}) = \sum_{k o_1 o_2 s s^\prime}
    \bar{\psi}_{k o_1 s} (-i\omega_n \delta_{s s^\prime} \delta_{o_1 o_2} + H^0_{o_1 o_2 s s^\prime} + \Delta_{o_1 o_2} \sigma_z^{s s^\prime}) \psi_{k o_2 s^\prime}
    - \frac{1}{U} \sum_o \Delta_{oo}^2 \,.
\end{equation}
With this effective action, the partition sum \cref{eq:free_energy} can integrated to yield
\begin{equation}
    F = \frac{1}{U} \sum_o \Delta_{oo}^2 - \frac{1}{\beta N} \sum_{kn} \ln(1 + e^{- \beta E_{kn}} ) + \mu n_\mathrm{tot} \,,
    \label{eq:free_energy_MF}
\end{equation}
with quasiparticle energies satisfying the eigenvalue equation
\begin{equation}
    \det\big( \underbrace{
        H^0_{o_1 o_2}(\bvec k) \delta_{s s^\prime} + \Delta_{o_1 o_2} \sigma_z^{s s^\prime}
    }_{ \mathcal H(\bvec k) } - E_{\bvec kn} \delta_{o_1 o_2} \delta_{s s^\prime} \big) = 0 \,.
    \label{eq:eigenvalue}
\end{equation}

Since the effective Hamiltonian is block diagonal in spin space, we exploit this symmetry: $\Delta_{BB} \sigma_z^{\uparrow \uparrow} = \Delta_{CC} \sigma_z^{\downarrow \downarrow}$. In conjunction with $C_{4v}$ symmetry of the free Hamiltonian ($H^0(\bvec k) = H^0_{B \leftrightarrow C}(C_4 \bvec k)$), it becomes evident that the contribution of the spin-$\uparrow$ and $\downarrow$ blocks to the free energy are identical. So we constrain $E_{\bvec k n}$ to the eigenvalues of the spin up block $H_{\uparrow \uparrow}(\bvec k)$ given by \cref{eq:order_parameter} and we can simply add a spin factor of $2$ to the corresponding term in the free energy.

The gap magnitude $\Delta_M$ is chosen such that the free energy \cref{eq:free_energy_MF} is minimized, i.e.,
\begin{equation}
    \frac{\partial F}{\partial \Delta_M} =
        \frac{4}{U} \Delta_M + \frac{2}{\beta N} \sum_{\bvec k n} \frac{1}{1 + e^{\beta E_{\bvec k n}}} \frac{\partial E_{\bvec k n}}{\partial \Delta_M} =
    \frac{4}{U} \Delta_M + \frac{2}{\beta N} \sum_{\bvec k n} f(\beta E_{\bvec k n}) \frac{\partial E_{\bvec k n}}{\partial \Delta_M} = 0 \,.
    \label{eq:gap_function}
\end{equation}
Applying the matrix identity ($\det(M)^{i i^\prime}$ denotes the minor determinant)
\begin{equation}
    \frac{\partial}{\partial x} \det(M) = \sum_{i i^\prime} \frac{\partial M_{i i^\prime}}{\partial x} (-1)^{i - i^\prime} \det(M)^{i i^\prime}
\end{equation}
to the eigenvalue problem \cref{eq:eigenvalue}, one can express the derivative of the eigenenergies as~\cite{Reiss2007}
\begin{equation}
    \frac{\partial E_{\bvec k n}}{\partial \Delta_M} =
    \frac{1}{\sum_o \det(M^n(\bvec k))^{o o}}
    \sum_{o_1 o_2} \frac{\partial \mathcal H_{o_1 o_2}(\bvec k)}{\partial \Delta_M} (-1)^{o_1 - o_2} \det(M^n(\bvec k))^{o_1 o_2} ~,
\end{equation}
where we defined the matrix $M^n_{o_1 o_2}(\bvec k) = \mathcal H_{o_1 o_2}(\bvec k) - E_{\bvec k n} \delta_{o_1 o_2}$. Using the shape of the order parameter and inserting the result in Eq.~\eqref{eq:gap_function}, this gives the final expression for the self-consistent gap equation
\begin{equation}
    \Delta_M = - \frac{U}{2 N} \sum_{\bvec k n} \frac{f(\beta E_{\bvec k n})}{\sum_o \det(M^n(\bvec k))^{o o}} \, \Big( \det(M^n(\bvec k))^{B B} - \det(M^n(\bvec k))^{C C} \Big) \,.
    \label{eq:gap_equation}
\end{equation}
The solutions for \cref{eq:gap_equation} depicted in \cref{fig:3} of the main text were obtained with a Brillouin zone sampling of $2000 \times 2000$ $\bvec k$ points.

\section{Long range magnetic order in 2D}
It is well known, that two dimensional spin systems lack true long range order due to the Mermin-Wagner-Hohenberg (MWH) theorem at any finite temperature~\cite{Mermin1966, Hohenberg1967}.
While it predicts a finite correlation length for the spin-spin expectation value, this length scale can be exponentially large.
Combined with the omnipresent coupling of any real two dimensional system with the three dimensional environment in physical conditions renders the implications of the MWH theorem irrelevant for practical use cases.

In the present theoretical model, that considers a real two dimensional material, we would nevertheless expect the MWH theorem to hold.
MF theory does not capture quantum fluctuation responsible for the destruction of long range coherence and is hence not expected to comply with the MWH theorem.
The FRG on the other hand takes into account fluctuations in an unbiased manner and still predicts a divergence of a collective spin excitation mode at finite temperature (or likewise critical cutoff) at odds with MWH.
This is a well known flaw of the 1-loop approximation of the FRG flow equations visualized in \cref{fig:diagrams} required to conveniently solve the infinite hierarchy of coupled differential equations~\cite{Defenu2015}.
It has been shown, that the multi-loop extension to the FRG is able to reproduce exact solutions in accordance with MWH~\cite{Kugler2018, Hille2020}.
However, these extensions are not applicable to multi-orbital systems due to their immense computational costs.
Still, the FRG in its 1-loop approximation is able to qualitatively predict the most dominant collective excitations, that will form quasi-long range order, that can be identified with a true long range coherence for all practical purposes even in the strict 2D limit.

\section{Stability analysis of the altermagnetic phase}
The results shown in the main part of this work were obtained for NN ($t=1$) and NNN hopping ($t'=0.5$), and equal onsite interactions and chemical potentials on the three sublattices ($U=1$, $\mu_i=0$) as given by the interaction Hamiltonian in \cref{eq:LiebHubbardModel} of the main text. The filling is adjusted such that the $t'$-induced VHS resides at zero energy. To investigate the resilience of the proposed AM state we consider perturbations in both the kinetic and the interaction part of the simplified Hamiltonian. We find the AM instability being robust against several changes summarized in the following subsections. Unless stated otherwise, the remaining parameters are kept as specified in this paragraph.

\subsection{Interaction parameters}
\begin{figure*}[t]
    \centering
    \includegraphics[width=0.9\linewidth]{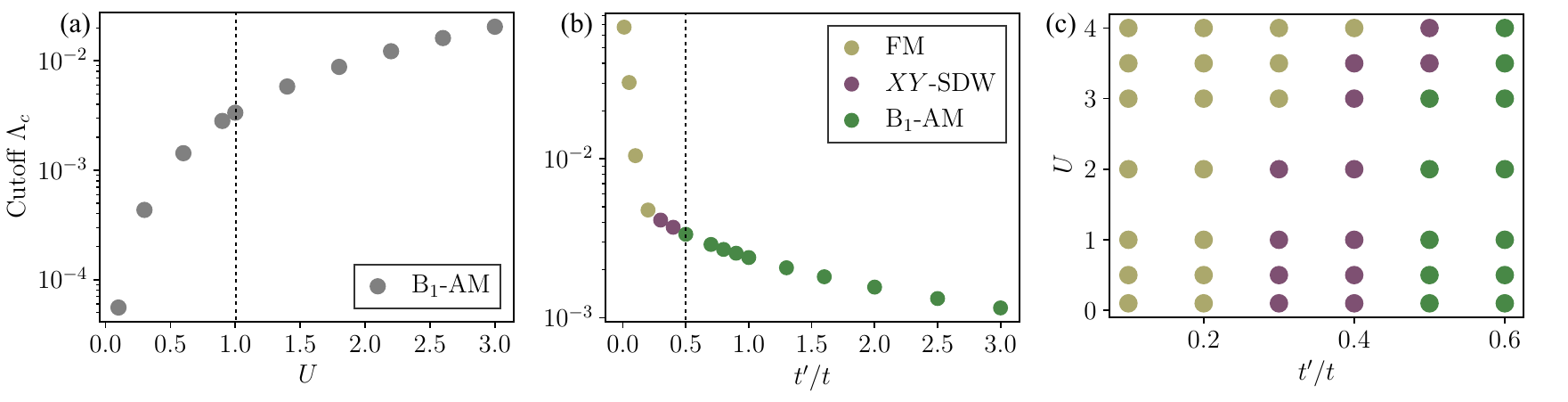}
    \caption{Critical cutoff scale $\Lambda_c$ of (a)~the proposed AM state for different values of $U_A = U_{B,C}$ at $t'=0.5$ and (b)~competing phases for different values of $t'$ at $U_i=1$ depicted in \cref{Fig:states}. (c)~Joint phase diagram the $U, t^\prime$ parameter space. We see that the AM state is stable in a broad parameter regime and extends to relatively large values of interaction.}
    \label{Fig:U_dep}
\end{figure*}

We demonstrate that the proposed AM state remains robust across the entire weak to intermediate coupling regime of onsite interactions $U_A = U_{B,C}$. \Cref{Fig:U_dep} a shows the critical cutoff scale $\Lambda_c$ for different values of $U$. The cutoff scale, being a handle on the critical temperature, increases rapidly with increased coupling strength. Furthermore, we analyze how different onsite interactions on the two types of sublattices affect the stability of the state, i.e. $U_A \neq U_{B,C}$. For fixed $U_{B,C}=1$ the AM state persists for $U_{A} = 0\, \dots \, 2.5$.

\begin{figure*}[t]
    \centering
    \includegraphics[width=0.75\linewidth]{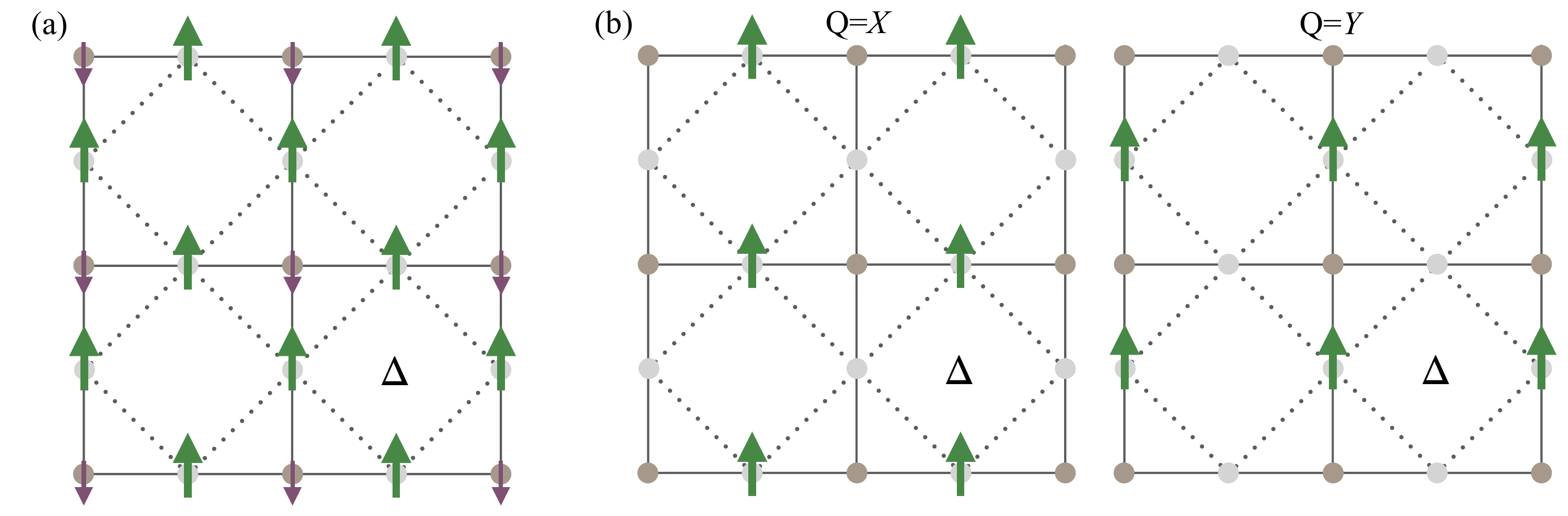}
    \caption{(a)~Ferromagnetic (FM) phase emerging at small $t'$. As the $A$ site participates in the magnetization process, a slight tendency towards ferrimagnetism can be observed. (b)~$XY$-spin density wave emerging at the interface of FM and AM. At the phase transition, all linear combinations of the twofold degenerate order parameter are allowed solutions of the linearized gap equation.}
    \label{Fig:states}
\end{figure*}

We additionally allow for longer-ranged interactions, i.e., nearest-neighbor (NN) interactions $V$ and next-nearest-neighbor (NNN) interactions $V'$. While NN interactions between Wykoff positions $1a$ and $2c$ have no effect on the AM state up to $V=0.6$, $V'$ (which couples the $B$ and  $C$ sites) drives a $B_1$ intra-unit cell charge density wave upon exceeding $V' = 0.2$. These results are in agreement with previous mean-field results on the Lieb lattice~\cite{Tsai2015} and are driven by an effect extensively discussed in Hubbard model on the Kagome lattice: Due to the sublattice interference mechanism~\cite{Kiesel2012} the system is less likely to form a local magnetization density, since the onsite interaction is partially screened by the sublattice character of states on the FS. Hence, the system has more incentive to ease the NN repulsion by a charge imbalance of the neighbouring sites rather than the onsite repulsion~\cite{Kiesel2013, Profe2024}.

\subsection{Kinetic parameters}

\begin{figure*}[t]
    \centering
    \includegraphics[width=0.9\linewidth]{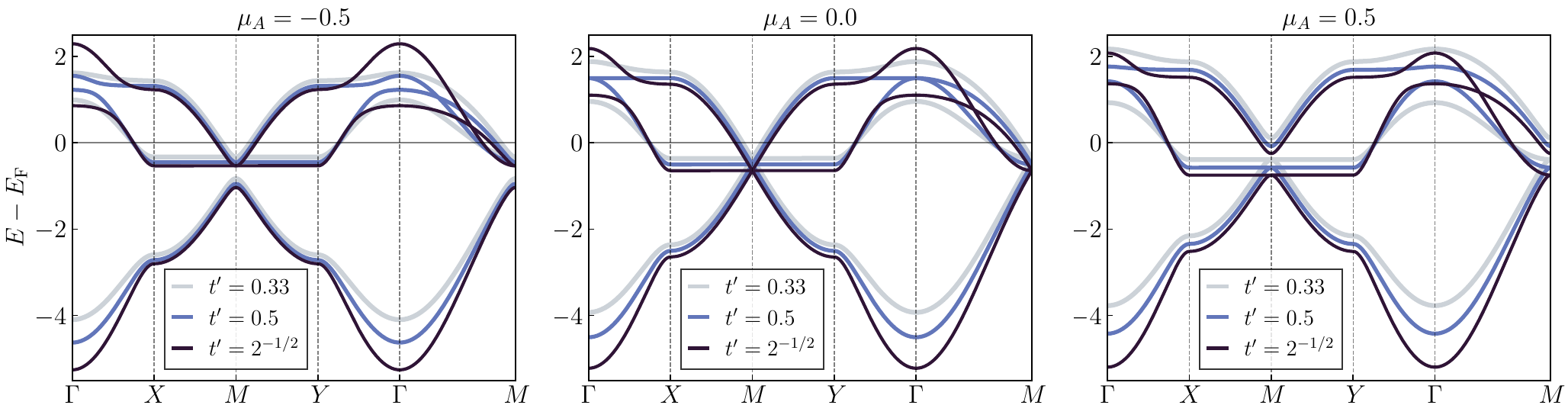}
    \caption{Band structures for different values of intrinsic detuning $\mu_A \neq \mu_{B,C}=0.0$ and NNN hopping $t'$ at pristine filling. The sublattice polarized VHS that drives the AM phase transition persits. For $t'\rightarrow 0$ the VHS approaches half filling and the band flattens.}
    \label{Fig:sup2}
\end{figure*}

\Cref{Fig:sup2} displays different values of intrinsic detuning $\mu_{A} \neq \mu_{B,C}=0.0$ and NNN hopping $t'$ at pristine filling. The sublattice polarized VHS at the $X$/$Y$ point persists for arbitrary values of $\mu_{A}$ and arbitrary but finite values of $t'$ as it corresponds to a topologically protected band touching point. This property is reflected in the resilience of the AM phase upon variation of $t'\geq 0.5$ (see \cref{Fig:U_dep}b) and $\mu_{A} = -2.0\dots3.0$.

At $U=1$ and small $t'<0.3$ the dominant contribution to the leading instability is provided by a ferromagnet (FM) on the $B$ and $C$ sites. The trivial irrep in real space allows for a participation of the central $A$ site in the magnetisation process, which results in a ferrimagnetic phase (\cref{Fig:states}a). A degenerate $Q=X,Y$ spin density wave (SDW) resides at the interface between FM and AM order (\cref{Fig:states}b). A phase diagram showing the three competing instabilites depending on both $U$ and $t'$ is shown in \cref{Fig:states}c.
The competition between these three states is already apparent at the bare level as discussed in \cref{sec:susceptibility}.
A sign difference between $t$ and $t'$ does not affect this phenomenology, as it merely inverts the dispersion of the BC band in energy.  
Small perturbations from even longer-ranged hybridizations, i.e., same-sublattice third nearest-neighbor hoppings $t''$, alter the shape of the Fermi surface (FS), resulting in a loss of perfect nesting conditions and, consequently, a suppression of critical scales.

For the given band structure, itinerant phases are generally suppressed for small amounts of hole doping as the DOS rapidly declines (see \cref{fig:1}b). For small amounts of electron doping, the $B$/$C$ polarized band is still present at the Fermi surface stabilizing the AM phase with tendencies to form the particle-hole condensate at slightly incommensurate momenta. As the logarithmically diverging DOS and perfect nesting scenario is absent at the Fermi level, intermediate interaction values are necessary to obtain the altermagnetic phase in our FRG calculations (e.g., $U_i = 1.4\dots\geq 2.6$ for $\mu = 0.02$).

\subsection{Dominance of the altermagnetic order parameter}
\label{sec:susceptibility}
To connect the orbital character of the eigenstates at the VHS with the emergent symmetry breaking, we inspect the bare static particle-hole susceptibility
\begin{equation}
   \chi^0_{o_1o_2o_3o_4}(\bvec{Q}) =
    - \int_\mathrm{BZ}\,\frac{\mathrm{d}\bvec k}{V_\mathrm{BZ}} \frac{f(\beta \varepsilon_n(\bvec k + \bvec Q)) - f(\beta \varepsilon_m(\bvec k))}{\varepsilon_n(\bvec k + \bvec Q) - \varepsilon_m(\bvec k)} M^{nm}_{\{o_i\}}(\bvec k, \bvec Q)
    \ .
    \label{eq:chi_0}
\end{equation}
The momentum space integral and Fermi distribution $f(\beta\epsilon)$ is evaluated on a momentum space mesh in the BZ of volume $V_\mathrm{BZ}$ and at an inverse temperature $\beta$ in the implementation described in Ref.~\cite{Duerrnagel2022}.

\begin{figure}[]
    \centering
    \includegraphics[width=0.6\columnwidth]{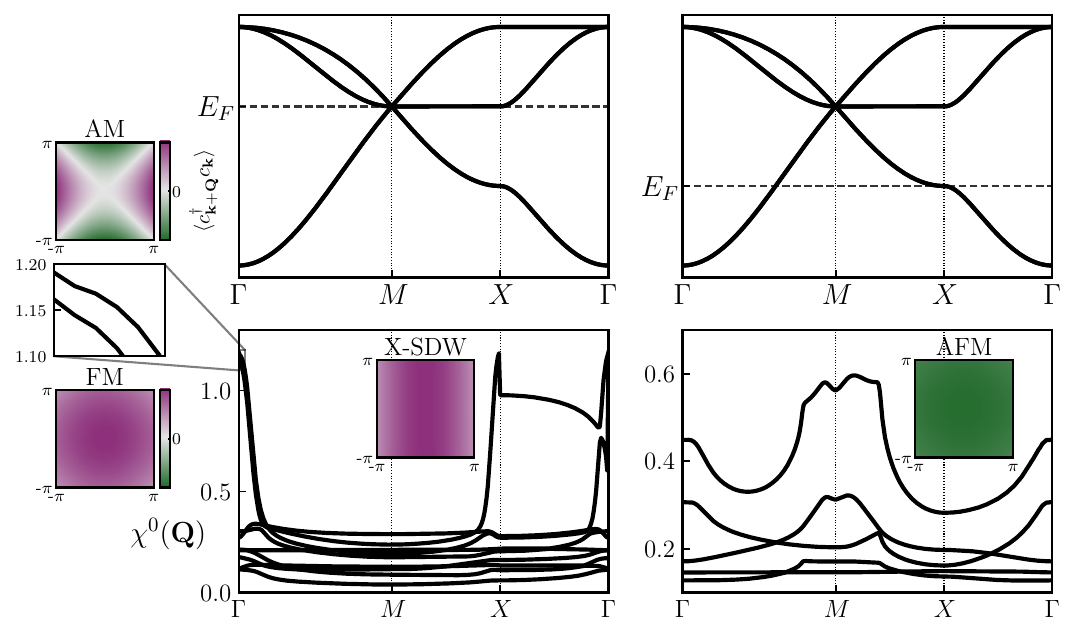}
    \caption{Bandstructure and eigenvalues of the bare susceptibility at lower VHS (left) and middle VHS (right) for the Hamiltonian in \cref{eq:LiebHubbardModel} of the main text for $t=-1$, $t^\prime = 0.5 t$ and $\mu_A = 0$. We evaluated the bare susceptibility via \cref{eq:chi_0} on a $\bvec{k}$-mesh of $2000 \times 2000$ and set $\beta = 100 / t$. The insets shows the eigenvector associated with the leading eigenvalues at distinct $\bvec{Q}$ values in the whole BZ. We note that ferromagnetic fluctuations are suppressed on the bare level already when doping to the center VHS (left column).}
    \label{Fig:sup3}
\end{figure}

In multi-orbital systems, $\chi^0$ features a dependence not only on the single particle energies $\varepsilon_n(\bvec k)$ but also on the orbital-to-band transformations of the electronic eigenstates $u^n_{o_i}(\bvec k)$ via
\begin{equation}
    M^{nm}_{\{o_i\}}(\bvec k, \bvec Q) =
    [u^n_{o_1}(\bvec k + \bvec Q)]^* 
    [u^m_{o_2}(\bvec k)]^* 
    u^n_{o_3}(\bvec k + \bvec Q) 
    u^m_{o_4}(\bvec k) \,.
\end{equation}
Since higher energy fluctuations are suppressed by the denominator in \cref{eq:chi_0}, the largest contribution to the integral is given by the states close to the Fermi level. This introduces an energetic hierarchy to the eigenstates on the bandstructure and one can select the dominant screening channels by placing the Fermi level accordingly. This effect is highlighted in \cref{Fig:sup3}: For the Fermi level at the VHS close to half filling, the bare susceptibility features a pronounced peak at $\Gamma$ and $X$ associated with the dominant FS nesting vectors on the $BC$ derived Fermi sheet. Contrarily, at the lower VHS the eigenstates on the FS are mostly dominated by the $A$ sublattice and the pronounced $M$ nesting displays an AFM spin fluctuation texture.

At the central VHS, the bare susceptibility shows two dominant peaks:
While at $\Gamma$ the susceptibility is guaranteed to feature a local maximum by quantum geometric arguments~\cite{kitamura2024spin}, the relative angular momentum of the resulting particle-hole condensate is determined by the ratio $t^\prime/t$.
NNN hopping ($t^\prime$) mediates a direct AFM coupling between the $B$ and $C$ sites $\propto t'^2$ that is counteracted by a second order FM coupling via the $A$ site of order $t^4$.
Their competition is directly reflected in the close proximity of the leading spin fluctuation in \cref{Fig:sup3} at the $\Gamma$ point.
In the $t^\prime$-dominated regime the associated electronic fluctuations display a clear $d$-wave altermagnetic character in accordance with the intuitive picture provided in the main text with a subleading FM order (see \cref{Fig:states}a).
For small $t^\prime/t$, this hierarchy is inverted resulting in the low energy FM phase in \cref{Fig:U_dep}b.

The additional peak at $\bvec Q = X/Y$ corresponds to the dominant Fermi surface nesting vector of the FS at the Lifshitz transition. It is generically disfavored at the bare level compared to the $\Gamma$ point condensate, since it can not exploit the mixed sublattice contributions directly at the van-Hove point $M$. However, the destructive interference of the AM and FM fluctuations at $\bvec Q = \Gamma$ promotes an $X/Y$ modulated spin density wave as the dominant magnetic instability in the intermediate $t^\prime$ region (cf.~\cref{Fig:U_dep}). 

The dominant peak in the bare susceptibility not only allows to identify the mot prominent ph fluctuations in the system, but directly transfers to the transition temperature in a general MF study of the system.
To see this, we can perform a generic mean field decomposition of the free energy in \cref{eq:free_energy} and expand it up to second order in the magnetic order parameters $\Vec{\Delta}$ to obtain the Ginzburg-Landau (GL) functional
\begin{equation}
    F_\text{GL} = \sum_\mathbf Q \Big[ \frac{1}{U} - \chi^0(\mathbf Q) \Big] \Vec{\Delta}(\mathbf Q)^2 + \mathcal{O}(\Vec{\Delta}^4) \ ,
\end{equation}
where we sum over different potentially degenerate wave vectors $\mathbf Q$ and dropped the orbital and spin indices of $\Vec{\Delta}$ and $\chi^0$ brevity.
This approximation is valid directly at the phase transition where $\Vec{\Delta} \rightarrow 0$ and requires more and more terms as one enters the ordered phase.
The expected transition temperature $T_c$ is given by the temperature for which the determinant of the second order coefficient matrix.
In the case of a single ordering vector, $T_c$ is determined by the condition $1/U = \chi^0(\mathbf Q)$, where the bare susceptibility is temperature dependent as evident from \cref{eq:chi_0} and grows as $T \rightarrow 0$.
Since we expect no change in the hierarchy of eigenstates as T is lowered, the eigenvalues of $\chi^0(\mathbf Q)$ in \cref{Fig:sup3} give a direct estimate for the free energy gain in the symmetry broken phase for $T < T_c$.
Consequently, we expect already at the MF level the AM state to provide the predominant instability for the flat band at the Fermi level.
It is also worth noting, that the first terms in the GL expansion, that allows a coupling of the different magnetic states present in \cref{Fig:U_dep} is of fourth order due to the different irreducible representations of the AM ($B_1$), FM ($A_1$) and $XY$-SDW ($\mathbf Q = X/Y$) order parameter
\begin{equation}
    F_\text{GL}^4 = \Vec{\Delta}_\text{AM}^2 \Vec{\Delta}_i (\mathbf Q_i)^2 + (\Vec{\Delta}_\text{AM} \cdot \Vec{\Delta}_i(\mathbf Q_i))^2 \,
\end{equation}
where $\Vec{\Delta}_i(\mathbf Q_i) \in \{ \Vec{\Delta}_\text{FM}(\mathbf Q = 0), \Vec{\Delta}_\text{SDW}(\mathbf Q = X/Y) \}$.
Therefore, we do not expect a sizable admixture of other magnetic states in the AM phase.
This validates the single component contrained MF calculation of \cref{sec:mean_field}.

\section{Symmetry classification of the $d$-wave altermagnetic order}
Non-relativistic collinear magnetism is generally classified by spin groups~\cite{Smejkal2022,Zeng2024} described as the direct product: $r_s\times R_s$. Here, $r_s$ refers to the spin only group whose implication on the band structure of collinear magnets is $\epsilon^{}_{\sigma}(k^{}_{x},k^{}_{y})=\epsilon^{}_{\sigma}(-k^{}_{x},-k^{}_{y})$ regardless of whether the real-space inversion symmetry is present.
Here, $r_s=[C_{\infty}||E]+[\bar{C}_2C_{\infty}||E]$ where $C_{\infty}$ is any rotation around the common spin axis and $\bar{C}_2$ is the two-fold rotation around the axis perpendicular to the spins combined with spin space inversion. In the notation $[R_i||R_j]$, $R_i(R_j)$ incorporates all the spin (crystallographic) space operations.

In contrast, the non-trivial spin group $R_s$, which does not contain any elements of $r_s$, plays a crucial role in determining the non-relativistic spin-split band structure. The group $R_s$ generally comprises pairs of operations acting independently on spin and real space. There are three distinct types of $R_s$, each corresponding to a specific type of collinear magnet.
\begin{enumerate}
\item \textbf{Type I}: Denoted by $R^{I}_s=[E||G]$ where $G$ belongs to a crystallographic Laue group. This type results in complete spin splitting of the band structure, as observed in ferromagnets or ferrimagnets.
 \item \textbf{Type II}: Denoted by $R^{II}_s=[E||G]+[C^{}_2||G]$ where $C^{}_2$ is a spin-inversion operation. This leads to a spin-degenerate band structure, typical of trivial collinear antiferromagnets.
 \item \textbf{Type III}: Denoted by $R^{III}_s=[E||H]+[C_2||G-H=A]$ where $H$  is a halving subgroup of $G$. This type gives rise to the altermagnetic phase, characterized by non-trivial non-relativistic spin splitting in spin-compensated collinear magnets.
\end{enumerate}
The observed phase in this study belongs to the third category. Specifically, the non-trivial spin group is given by $[C_2||A]$ where $A$ includes three symmetry elements $\mathcal{M}^{}_{xy}$ (mirror plane perpendicular to $xy$ plane and passing through $x=y$), $\mathcal{M}^{}_{x\bar{y}}$ (mirror plane perpendicular to $xy$ plane and passing through $x=-y$) and $C^{}_4$ ($C^{}_{4z}$ rotational axis) that interchanges B and C sublattices:
\begin{equation}
\begin{aligned}
        [C^{}_2||\mathcal{M}^{}_{xy}]:\;&\epsilon^{}_{\sigma}(k^{}_x,k^{}_y)\rightarrow\epsilon^{}_{-\sigma}(k^{}_y,k^{}_x) \,, \\
        [C^{}_2||\mathcal{M}^{}_{x\bar{y}}]:\;&\epsilon^{}_{\sigma}(k^{}_x,k^{}_y)\rightarrow\epsilon^{}_{-\sigma}(-k^{}_y,-k^{}_x) \,, \\
        [C^{}_2||C^{}_{4}]:\;&\epsilon^{}_{\sigma}(k^{}_x,k^{}_y)\rightarrow\epsilon^{}_{-\sigma}(-k^{}_y,k^{}_x) \,.
\end{aligned}
\end{equation}
The first two equation ensure the symmetry protected spin degeneracy along the lines $k^{}_{x}=\pm k^{}_{y}$, i.e., along the segment $\Gamma$-$M$.

\section{Size of the non-relativistic spin splitting}
In the magnetic state, spin degeneracy is lifted and the overall Hamiltonian can be written in a block diagonal fashion
\begin{equation}
    \mathcal H(\bvec k) =
    \begin{pmatrix}
        \mathcal H_{\uparrow \uparrow}(\bvec k) & 0 \\
        0 & \mathcal H_{\downarrow \downarrow}(\bvec k) \ .
    \end{pmatrix}
\end{equation}
Even in the presence of magnetic ordering the two spin block decouple and we can obtain the eigenvalues of the spin-$\uparrow$ polarized bands by diagonalization of the matrix
\begin{equation}
    \mathcal H_{\uparrow\uparrow}(\bvec k) =
    \begin{pmatrix}
        - \mu_A & A_{k_x} & A_{k_y} \\
        A_{k_x} & \Delta_M & B_k \\
        A_{k_y} & B_k & -\Delta_M
    \end{pmatrix}
\label{eq:AM_hamiltonian}
\end{equation}
with $\Delta_M$ ($-\Delta_M$) the spin polarization on the $B$ ($C$) sublattice.
The corresponding eigenvalues for spin-$\downarrow$ are obtained by flipping the sign: $\Delta_M \rightarrow -\Delta_M$.

\begin{figure*}[t]
    \centering
    \includegraphics[width=0.25\linewidth]{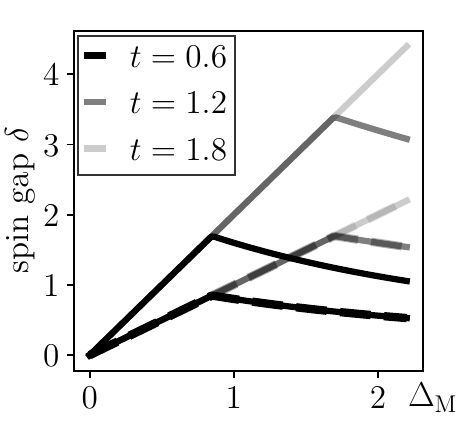}    \caption{Magnitude of the spin gap $\delta$ as a function of the magnetic order parameter magnitude $\Delta_M$ for several values of NN hopping $t$ with $t'=t/2$. The kinks correspond to level crossings of the altermagnetic bands, so the linear scaling abruptly changes there. }
    \label{Fig:spingap}
\end{figure*}

To assess the size of the spin splitting we recall that the magnetic order parameter transforms in a $B_1$ irrep of $C_{4v}$, i.e., the spin splitting is to lowest order $\propto \cos(k_x) - \cos(k_y)$~\cite{Roig2024}. Hence, the maximum value of the spin splitting is expected at the $X/Y$ points. We therefore investigate the Hamiltonian at $X = (\pi, 0)$,
\begin{equation}
    \mathcal H_{\uparrow\uparrow}(\bvec k = X) =
    \begin{pmatrix}
        - \mu_A & 0 & 2t \\
        0 & \Delta_M & 0 \\
        2t & 0 & -\Delta_M
    \end{pmatrix} \,,
\end{equation}
to quantify the spin splitting. The eigenvalues of this matrix as well as its spin-down counterpart are given by
\begin{equation}
\begin{aligned}
    E^{1}_\uparrow(X) &{}= -\Delta_M \,, &
    E^{1}_\downarrow(X) &{}= \Delta_M \,,\\
    E^{2}_\uparrow(X) &{}= \frac{1}{2} \left( \Delta_M + \mu_A - \sqrt{(\Delta_M - \mu_A)^2 + 16 t^2} \right) \,, &
    E^{2}_\downarrow(X) &{}= \frac{1}{2} \left(-\Delta_M + \mu_A - \sqrt{(\Delta_M + \mu_A)^2 + 16 t^2} \right) \,, \\
    E^{3}_\uparrow(X) &{}= \frac{1}{2} \left( \Delta_M + \mu_A + \sqrt{(\Delta_M - \mu_A)^2 + 16 t^2} \right) \,, &
    E^{3}_\downarrow(X) &{}= \frac{1}{2} \left(-\Delta_M + \mu_A + \sqrt{(\Delta_M + \mu_A)^2 + 16 t^2} \right) \,,
\end{aligned}
\end{equation}
where we require the labelling with $i$ such that $E^i_\uparrow(X)= E^i_\downarrow(X)$ in the non-magnetic case ($\Delta_M = 0$).
Then the non-relativistic spin splitting at the $\bvec k=X$ point is given by $\delta^i(X) = |E^i_\uparrow(X) - E^i_\downarrow(X)|$ (for $\Delta_M \ll 1$, when the bands stay close to the non-magnetic bands) and reads
\begin{equation}
\begin{aligned}
    \delta^1(X) &{}= 2 \Delta_M \,, \\
    \delta^2(X) &{}= \Big| \Delta_M - \sqrt{(\Delta_M - \mu_A)^2 + 16 t^2} +
                  \sqrt{(\Delta_M + \mu_A)^2 + 16 t^2} \Big| 
                  = \Delta_M \Big| 1 + \frac{\mu_A}{t} + \mathcal O(t^{-2})\Big| \,, \\
    \delta^3(X) &{}= \Big| \Delta_M + \sqrt{(\Delta_M - \mu_A)^2 + 16 t^2} -
                 \sqrt{(\Delta_M + \mu_A)^2 + 16 t^2} \Big|
                  = \Delta_M \Big| 1 - \frac{\mu_A}{t} + \mathcal O(t^{-2}) \Big| \,. 
\end{aligned}
\end{equation}
Hence the spin gap $\delta^i(X)$ scales linearly in $\Delta_M$ for $\Delta_M \ll t$ in accordance with an AFM band gap. As depicted in \cref{Fig:spingap} this linear regime extends over a broad range of order parameter sizes until a level crossing of the spin polarized bands sets in.

A non-itinerant mechanism to generate altermagnetic order usually relies on strong magnetic exchange interactions. They are hence operated in a regime where the magnetic order is much larger than the local symmetry breaking term. In this limit, corresponding to $t \ll \Delta_M$, the spin gap of our system is given with reference to the AFM state in the BC subsystem by $\bar \delta(X) \propto t^2/|\Delta_M - \mu_A|$.
Hence, the AM spin splitting is determined by the effective hopping term that transforms non-trivially under $C_4$ (cf.~\cref{sec:perturbation_theory}) in accordance with the effective AM toy models discussed, e.g., in Ref.~\cite{Smejkal2022}. The transition between an ``itinerant gap''{} and a ``local gap''{} can be seen in \cref{Fig:spingap}, where the kink at $\Delta_M\approx 1$ corresponds to the change of reference state.

\section{Realization in an optical lattice}

\begin{figure*}[]
    \centering
    \includegraphics[width=1\linewidth]{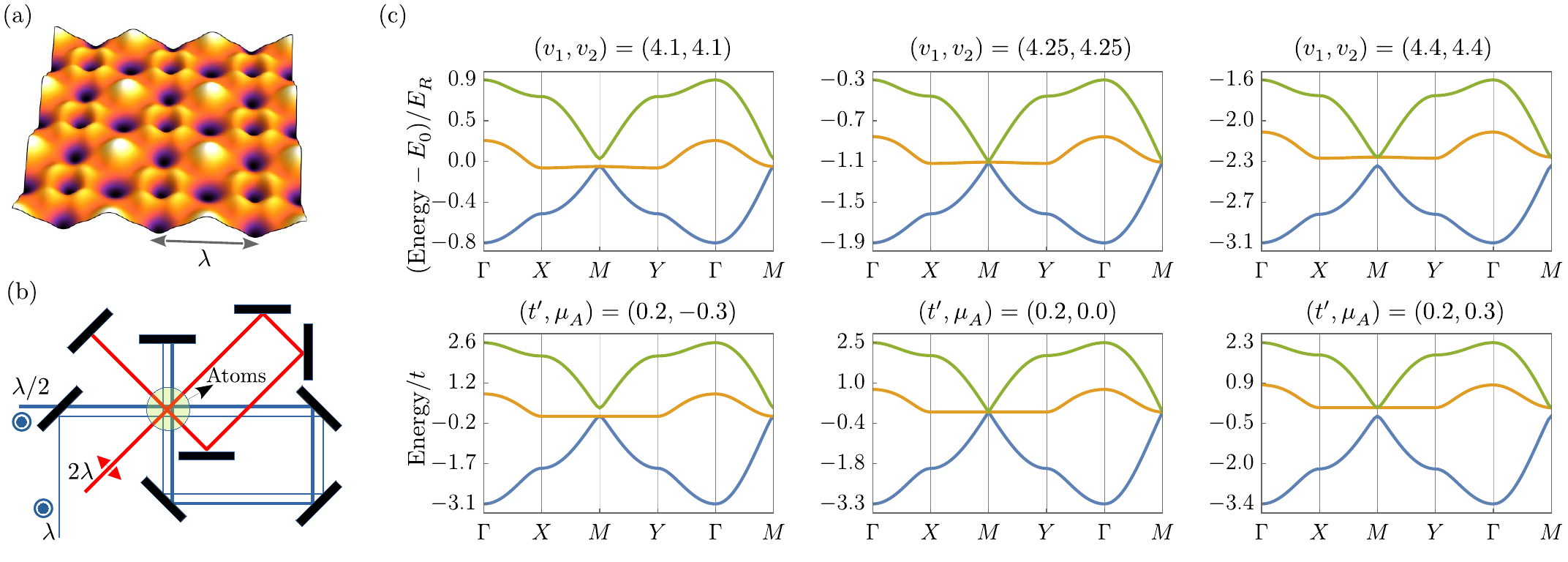}
    \caption{Possibility of realization Lieb system in an optical lattice simulator. (a)~Optical lattice potential given by \cref{eq:optical_potential} with $(v^{}_{1},v^{}_{2},r)=(4.0,4.0,0.7)$, (b)~Experimental setup. (c)~top panels: optical lattice band structure (single particle) for $(v^{}_{1},v^{}_{2})$ given by $(4.1,4.1)$, $(4.25,4.25)$ and $(4.1,4.1)$ respectively and $r=0.7$. We consider the energy offset as $E_0=26E_R$. (c)~bottom panels: Tight binding band structure with $(t',\mu^{}_{A})$ given by $(0.2,-0.3)$,$(0.2,0.0)$ and $(0.2,0.3)$ in \cref{eq:LiebHubbardModel} where $\mu^{}_A$ and $t'$ are scaled with $t$.}
    \label{Fig:optical_lattice_lieb}
\end{figure*}

The altermagnetic Hubbard model on the Lieb lattice consists of NN hopping $t$ and NNN hopping $t'$. In this section, we discuss a possible realization of the single particle band structure of such a model in an optical lattice platform. We begin with the following lattice potential:
\begin{align}
V(x,y)&=-E^{}_R[V^{}_1(x,y)+V^{}_2(x,y)],\label{eq:optical_potential}\\
V^{}_1(x,y)&=v^{}_{1}[\cos(2\kappa x)+\cos(2\kappa y)+r(\cos(4\kappa x)+\cos(4\kappa y))]^2,\\
V^{}_2(x,y)&=4v^{}_{2}[\sin^2(\kappa(x+y))+\sin^2(\kappa(x-y))],
\end{align}
where $E_R=\hbar^2\kappa^2/2m$ is the recoil energy, $m$ is the mass of ultra-cold atoms trapped into the potential wells and $\kappa=\pi/\lambda$. \Cref{Fig:optical_lattice_lieb}a illustrates this lattice potential for $(v_1,v_2,r)=(4.0,4.0,0.7)$.

In \cref{Fig:optical_lattice_lieb}b, we sketch the experimental setup yielding the potential of the form given by \cref{eq:optical_potential}. $V_1(x,y)$ can be generated by two laser sources that result in two pairs of phase-locked counter-propagating beams with wavelengths $\lambda/2$ and $\lambda$ shown by thick and thin blue lines respectively in \cref{Fig:optical_lattice_lieb}b. Both beams are linearly polarized with the direction of polarization aligned out of plane. Here, $r=\frac{|\Vec{E}^{}_{0}(\lambda/2)|}{|\Vec{E}^{}_{0}(\lambda)|}$ denotes the ratio of the electric field amplitudes of the two beams. On the other hand, to generate $V_2(x,y)$ one needs a linearly polarized single laser source yielding a pair of phase locked counter propagating waves of wavelength $2\lambda$ where the plane of polarization resides within the plane (shown by red lines in \cref{Fig:optical_lattice_lieb}b). Notice that this beam is rotated by $45^\circ$ with respect to the previous one. Such arrangement leads to a Lieb lattice with lattice periodicity $\lambda$.

Furthermore we follow standard techniques for calculation of the single particle band structure in a lattice potential. This includes numerically solving Schrödinger's equation with potential $V(x,y)$. As a result, we obtain band structures displayed in the top panels of \cref{Fig:optical_lattice_lieb}c for different parameter choices. A comparison of these spectra with the single particle band structures obtained from the model Hamiltonian (cf.~\cref{eq:LiebHubbardModel} of the main text) for corresponding values of the tight binding parameters is given in the bottom panels of \cref{Fig:optical_lattice_lieb}c. The good agreement suggests a tangible realization of the itinerant AM phase transition in optical lattice systems. In addition, there are other proposals~\cite{Liberto-2016,Shen-2010} suitable for the realization of an NN Lieb lattice tight-binding model.

\end{document}